# The Source of Hydrogen in Earth's Building Blocks

Thomas J. Barrett[1], James F.J. Bryson[1], Kalotina Geraki[2]

1. Department of Earth Sciences, University of Oxford, Oxford, OX1 3AN, UK.
2. Diamond Light Source, Harwell Science and Innovation Campus, Didcot, OX11 0DE, UK.

**Email**: thomas.barrett@st-annes.ox.ac.uk

## Abstract

Despite being pivotal to the habitability of our planet, the process by which Earth gained its present-day hydrogen budget is unclear. Due to their isotopic similarity to terrestrial rocks across a range of elements, the meteorite group that is thought to best represent Earth's building blocks is the enstatite chondrites (ECs). Because of ECs' nominally anhydrous mineralogy, these building blocks have long been presumed to have supplied negligible hydrogen to the proto-Earth. However, recent bulk compositional measurements suggest that ECs may unexpectedly contain enough hydrogen to readily explain Earth's present-day water abundance. Together, these contradictory findings mean the contribution of ECs to Earth's hydrogen budget is currently unclear. As such, it is uncertain whether appreciable hydrogen is a systematic outcome of Earth's formation. Here, we explore the amount of hydrogen in ECs as well as the phase that may carry this element using sulfur X-ray absorption near edge structure (S-XANES) spectroscopy. We find that hydrogen bonded to sulfur is prevalent throughout the meteorite, with fine matrix containing on average almost 10 times more H-S than chondrule mesostasis. Moreover, the concentration of the H-S bond is linked to the abundance of micrometre-scale pyrrhotite ($Fe_{1-x}S$, $0<x<0.125$). This sulfide can sacrificially catalyse a reaction with $H_2$ from the disk at high temperatures to create $H_2S$, which could be dissolved in adjoining molten silicate-rich material. Upon rapid cooling, this assemblage would form pyrrhotite encased in submicron silicate-rich glass that carries trapped $H_2S$. These findings indicate that hydrogen is present in ECs in higher concentrations than previously considered and could suggest that this element may have a systematic, rather than stochastic, origin on our planet.

**Key Words**
Meteorites, Planetary formation, Cosmochemistry

## Introduction

Although water is central to planetary habitability as we know it, the mechanism by which Earth gained its substantial present-day water budget is unclear. Enstatite chondrites (ECs) closely resemble terrestrial rocks across a range of isotopic compositions (Dauphas, 2017). Hence, ECs (or material similar to these meteorites) have been suggested as the predominant building blocks of the proto-Earth (Javoy et al., 2010), as well as other terrestrial planets including Mars (Brasser et al., 2018). ECs formed in the hot inner solar system, and consist entirely of nominally anhydrous minerals (Weisberg & Kimura, 2012). As such, Earth's building blocks have long been presumed to have been 'dry', and so provided negligible water to the proto-Earth (e.g., Marty, 2012; Morbidelli et al., 2000). This concept has recently been reinforced by measurements of very low water concentrations in nominally anhydrous silicates in aubrites (achondrites isotopically resembling ECs) (Peterson et al., 2023). Although it is unclear what proportion of hydrogen (and other volatile elements) would survive processes such as planetary degassing and differentiation, these measurements could imply that enstatite chondrites were effectively devoid of water.

Several processes have been suggested to have then provided water to our planet later in its history. These include a late veneer of hydrated cometary material (Owen & Bar-Nun, 1995), and delivery via bombardment from outer-solar-system derived carbonaceous chondrite-like material (Alexander et al., 2012). These models imply that our planet's water budget – and by extension, its suitability for life – is the result of the stochastic scattering of outer solar system material into the terrestrial planet forming region (Morbidelli et al., 2000). In this case, habitability would not be a natural consequence of Earth's formation from EC-like material.

Contrarily, recent measurements of the bulk compositions of ECs reveal they may unexpectedly contain high hydrogen concentrations (0.08 – 0.54 wt% $H_2O$)(Piani et al., 2020; Thomassin et al., 2023). Indeed, these measurements suggest ECs seemingly carry enough of this ingredient of water to account for up to ~14 times the mass of Earth's oceans. Moreover, isotopic measurements argue that Earth may have accreted as much as ~70 % of its volatile inventory, including hydrogen and nitrogen, from inner solar system material (Martins et al., 2023; Piani et al., 2020; Savage et al., 2022; Steller et al., 2022). Later delivery of outer solar system material then likely contributed small amounts of volatile elements, which help to resolve the observed hydrogen and nitrogen isotopic ratios of Earth's surface (Alexander et al., 2012; Piani et al., 2020). Combined, these findings suggest that Earth, and other terrestrial planets, may have in fact formed with high initial inventories of hydrogen and other volatiles. However, one pivotal unknown in these recent measurements is the phase that carries ~80 % of the bulk hydrogen concentration in ECs. Without this information, our understanding of the reason this element may be present in these meteorites in such high concentrations is severely limited.

In all, current observations disagree on the hydrogen content of ECs. As such, it is unclear whether the proto-Earth contained abundant hydrogen from its inception or whether this element was predominantly delivered to our planet after the main stage of its formation. It is therefore currently unknown whether Earth's water is a natural outcome of its formation from EC-like material. To explore the amount of native hydrogen in ECs as well as examine the phase that may host this element in these meteorites, here we map an EC using micrometre-scale X-ray absorption near edge structure spectroscopy at the sulfur K edge (S-XANES).

**Materials and Methods**

We explored the speciation of S, and the abundance of different species of this element, in the EH3 chondrite LAR 12252 using micrometre-scale S-XANES at beamline I18 at the Diamond Light Source. We chose to study LAR 12252 due to its low weathering grade of A and a description of only modest terrestrial alteration. This low amount of weathering minimises the likelihood of widespread and/or prevalent terrestrial H-bearing weathering products.

We measured a 2D S-XANES map of an area of clastic matrix in this meteorite. This region included fragments of chondrules (which included mesostasis), enstatite grains, metal grains, various sulfide grains, and fine matrix (composed of an assortment of materials, all <5 μm). We measured these maps by sweeping an X-ray beam with a 5 μm spot size in 5 μm steps over an area 470 × 480 $\mu m^2$ large and measuring an S-XANES spectrum at each step (resulting in a total of 9,024 spectra). The energy of the beam was increased from 2465-2495 eV in 0.2 eV intervals at each individual spot, with an acquisition time of 0.05 s at each energy value. The mapped area was selected to minimise the number of large objects (≳100 μm) and visible signs of terrestrial weathering, such as rust. We also measured 15 spot spectra at points within mesostasis in 6 chondrules throughout our thin section, as well as in multiple cracks containing terrestrial weathering products. Chondrule mesostasis spots were selected following optical microscopy, ensuring that they represented the range of chondrule sizes and types present in the thin section.

To analyse our data, we fitted all our S-XANES spectra with multiple Gaussian curves using IGOR Pro, discussed further in the SI. We chose this method over linear combination fitting due to the heterogeneities within the sample, absence of a complete reference database of spectra collected from meteorites, and the number of spectra collected. We initially fitted each spectrum only between 2465-2478 eV (Fig. 1b) because this energy range contains the iron sulfide peak, the H-S peak, and parts of the S-edge step function and peaks due to S species at higher energies. We chose this approach because it allowed us to focus on the peaks of interest, making the fitting procedure more robust and straightforward (see SI). As seen in Figure 1, this method produced only small fitting errors (discussed further in the SI). To determine the amplitude of the $S^{6+}$ peak, we then fitted a single Gaussian curve in the energy range 2580-2590 eV. This approach means the recovered height of this peak is that above the background in this energy range, so it isolates the $S^{6+}$ signal by not considering the S-edge step function or any other S species. To explore the amount of H-S throughout our mapped area of LAR 12252, we calculated the enrichment/depletion of H-S at each point in mapped area by normalising the amplitude of the H-S peak recovered from each spectrum to the average amplitude of this peak recovered from the 15 spot spectra of chondrule mesostasis. Chondrule mesostasis contains a heterogenous H concentration with a fairly wide range of measured values (Piani et al., 2020; Thomassin et al., 2023). This natural spread in concentration reflected itself in our study as a wide range of H-S peak amplitudes in mesostasis, which introduced the largest source of error in this study (discussed further in the SI). We isolated spectra that correspond to fine matrix using peak amplitude ratios (discussed further in the SI).

## Results

### *Peak Identification*

The S-XANES spectra we collected display a variety of peaks, the presence of which depends principally on the phase that was measured (Figure 1a). S-XANES spectra of fine matrix and mesostasis typically contain four peaks (Figure 1a), providing evidence of sulphur in different compounds and oxidation states in these phases. By comparing peak energy and shape to reference spectra (e.g., Anzures et al., 2019; Lerner et al., 2021; Prange et al., 2002) as well as from the Database of Inorganic Sulfur Compounds from The European Synchrotron Radiation Facility, we can identify the species responsible for each peak.

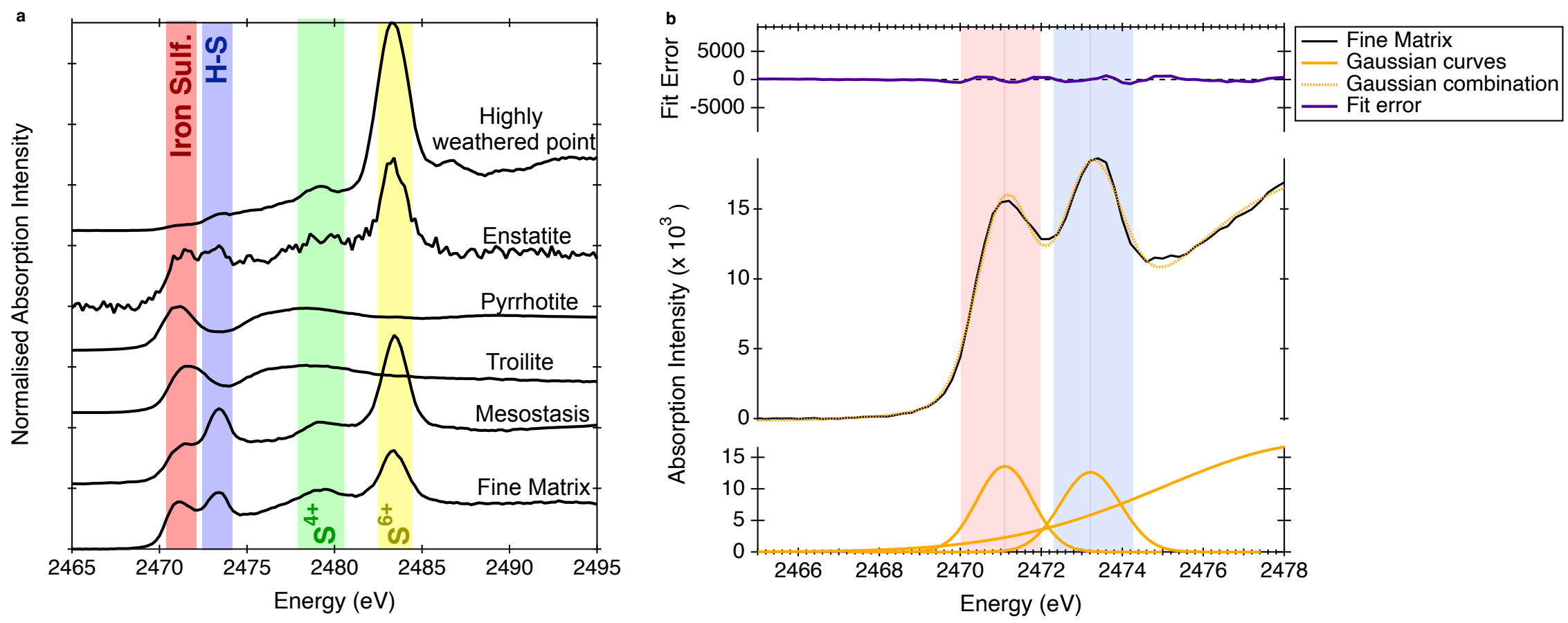

**Figure 1a** Representative S-XANES spectra measured in a variety of different phases in LAR 12252 during this study. Positions of peaks ascribed to species from reference spectra *(e.g., Anzures et al., 2019; Lerner et al., 2021; Prange et al., 2002)* are highlighted by vertical coloured bars. Spectra are normalised, so amplitudes here do not reflect sulfur content. Spectra from enstatite grains appear noisy due to very low sulfur counts, and closely resemble spectra measured in metal grains. The sulfur signal in these non-sulfur bearing phases are likely caused by minor sampling of matrix that could be adjacent to, or underneath these phases, or the presence of small amounts of sulfur-bearing impurities within these phases. **1b** Peak fitting method used to extract the amplitudes of the iron sulfide and H-S peaks from the fine matrix spectra in 1a. We achieved this multi-peak approach by creating a total fit (dashed orange in middle panel) from three Gaussian curves (orange, bottom panel) across the energy range 2468-2478 eV. This process yields small residual errors (purple, top panel). The dashed line in the top panel marks zero, and the total variation in the scale is set as the same as that in the central panel. The highlighted peaks correspond to the same as in Figure 1a.

The peak at lowest energy, highlighted in red in Figure 1, is widespread in fine matrix and mesostasis and is ascribed to submicron iron sulfides (troilite and pyrrhotite) (Anzures et al., 2019; Fleet, 2005; Klimm et al., 2012; Métrich et al., 2009). The energy of this peak varies from 2471.8 eV to 2471.2 eV, diagnostic of the composition of this mineral shifting from stoichiometric troilite to pyrrhotite ($Fe_{1-x}S$, $0<x<0.125$) (Head et al., 2018). Energies of all peaks detected in our spectra are slightly (~1 eV) higher than reported in most previous studies (e.g., Anzures et al., 2019; Fleet, 2005; Klimm et al., 2012; Métrich et al., 2009), but remained constant throughout the duration of our study. Such variations are common across XANES studies and are attributed to differences in calibration of the monochromator (see SI).

Throughout the fine matrix and mesostasis, we also find a distinctive peak at ~2473.2 eV, highlighted in blue in Figure 1. Due to similar peak locations in reference spectra, this peak could be due to hydrogen bonded to S, native S, or pyrite (Bose et al., 2024). However, based on previously reported spectra of $H_2S$ trapped in silicate glasses, gaseous $H_2S$, and thiol groups (Bodeur & Esteva, 1985; Bose et al., 2024; Klimm et al., 2012; Orthous-Daunay et al., 2010; Prange et al., 2002), the peak location relative to those of FeS and $S^{6+}$, and the absence of elemental S in previous sub-micron-scale compositional measurements (Lehner et al., 2014), we ascribe this peak to S bonded to H. Additionally, if this peak was caused by elemental S, rather than H-S, we would expect a broadening/shifting to higher energy of the wide absorption peak highlighted in green in Fig 1 than we observe. Our observation of H-S in chondrule mesostasis supports previous Raman spectroscopy measurements of the H-S bond in this phase (Thomassin et al., 2023). Additionally, our finding of the H-S peak throughout fine matrix demonstrates that H-S is a common form of hydrogen in ECs, and that this form of hydrogen is present in multiple phases.

In addition to reduced sulphides, many spectra also contain a peak at ~2483.4 eV corresponding to $S^{6+}$, the most oxidised form of sulfur, highlighted in yellow in Figure 1. The amplitudes of these peaks vary widely among the points we measured, and they are at their highest in spot spectra collected within cracks where clear terrestrial weathering has occurred (Fig. S3). They also exist at much lower, but still variable, amplitudes in matrix and mesostasis. Different sulfates can be difficult to distinguish using S-XANES because different cations do not cause large shifts in peak energy (Anzures et al., 2019). Despite this, we were able to identify $CaSO_4$ and $FeSO_4$ at some points in our mapped area of matrix, owing to minor peaks in the pre- and post-edge regions of the spectra (Anzures et al., 2019). The $S^{6+}$ peak is absent within macroscopic (>10 μm) grains of metal, enstatite, and sulfides.

Finally, we observe a small peak centred on ~2478 eV, shown in green in figure 1. The amplitude of this peak is affected by two species: a second, wide absorption feature of troilite and pyrrhotite; and small amounts of $S^{4+}$ likely formed through gradual photo-reduction of $S^{6+}$ by the X-ray beam (Lerner et al., 2021; Wilke et al., 2008). We minimised the impact of photoreduction by mapping with as short acquisition times as possible.

*Hydrogen Abundance*

Because peak amplitude correlates with species abundance (Anzures et al., 2019; Fleet, 2005; Klimm et al., 2012), we are able to explore concentrations of the different species in the matrix of LAR 12252 using S-XANES spectra. However, the nature of complex geological samples like meteorites (e.g., the wide range of measured H concentrations in mesostasis) introduces specific challenges that make absolute concentrations difficult to determine reliably via comparisons to datasets collected using other techniques (e.g., Piani et al., 2020; Thomassin et al., 2023). However, we were able to estimate the relative concentration of H-S in EC matrix by comparing the amplitude of the peak at ~2473.2 eV throughout our map to the average amplitude of this peak we measure in chondrule mesostasis. We found that our clastic matrix region contains points with H-S peak amplitudes that vary from 0 - ~40 times the average mesostasis amplitude (Figure 2b). The lowest relative amplitudes are found within the interiors of macroscale enstatite, metal, and sulfide grains. The highest amplitudes are in regions of fine matrix. Excluding points with more extensive terrestrial weathering (in the form of large $S^{6+}$ peaks, see *Discussion* section; Fig S5), we calculate that fine matrix in LAR 12252 contains an average of 9.8 ± 3.1 times as much H-S as chondrule mesostasis in this meteorite (2 standard error, calculated from the measured mesostasis H-S peak amplitudes; discussed further in SI).

By combining this value with the proportion of fine matrix in EH3 chondrites, we are able to constrain the relative contribution of fine matrix towards the H budget of these meteorites. Unfortunately, the parameters used to calculate the proportion of fine matrix have not been measured for LAR 12252, so we chose to use parameters measured from the paired EH3 chondrites SAH 97072 and SAH 97096. We adopted this approach because these meteorites are particularly well-studied, their bulk and mesostasis H concentrations have been measured previously (Piani et al., 2020), and the proportion of fine matrix is likely similar among most EH3 chondrites (±2 vol%; Lehner et al., 2014). As such, the value we calculate is likely relevant to the enrichment factor we recovered from LAR 12252. Using grain (3730 kg $m^{-3}$) and bulk (3240 kg $m^{-3}$) densities (Macke et al., 2010), the volume fraction of clastic matrix (21.5 vol%; Lehner et al., 2014), and the proportion of fine matrix in clastic matrix (~50 vol%) in these meteorites (Rubin et al., 2009), we calculate a mass fraction of fine matrix (Bryson & Brennecka, 2021) of ~4.8 wt%. This value is approximately half that of chondrule mesostasis in the same meteorites (9.3 wt%; Piani et al., 2020). As such, assuming fine matrix in these meteorites also carries on average ~9.8 times as much H as mesostasis, fine matrix would contribute an average of ~5 times more H to the bulk H concentration than mesostasis.

Using this factor, we can explore whether this H-S enrichment in fine matrix corresponds to a significant amount of H. For reference, previous studies have found that mesostasis accounts for an average of ~13% of the bulk H concentration in SAH 97096 (Piani et al., 2020). Because they have not been measured previously, it is unknown whether other EH3 chondrites share equally high mesostasis H concentrations. However, combining our recovered enrichment factor with this proportion of H in mesostasis argues that fine matrix likely contributes a significant amount of H towards the bulk concentration in EH3 chondrites (several tens of wt%), and could account for the majority of H in these meteorites.

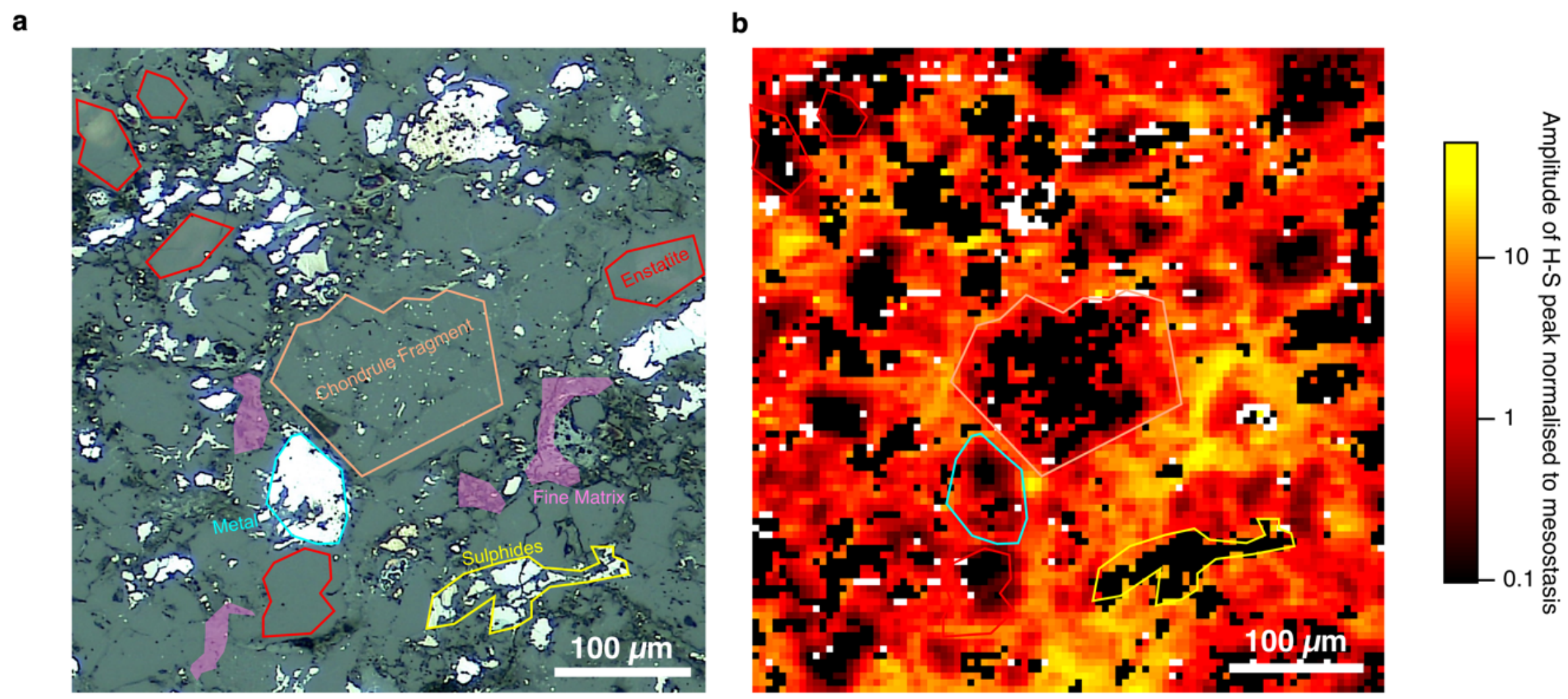

**Figure 2a** Optical microscopy image of the clastic matrix region in LAR 12252 that was studied using S-XANES. Notable phases include enstatite (examples outlined in red), sulphides (example outlined in yellow), a small chondrule fragment (outlined in orange), and fine matrix (examples highlighted in pink). **b** Corresponding map calculated from the S-XANES spectra of the region shown in **a**, showing the amplitude of the H-S peak at each point normalised to the average amplitude of point spectra measured from chondrule mesostasis (Fig. S8). Black pixels correspond to points where the S-XANES spectra that did not contain a resolvable H-S peak. White pixels correspond to points where the S-XANES spectra could not be fitted due to their quality or noise in the spectra. Some grain outlines have been transferred onto Fig 2b to allow for comparison of the two figures.

## Discussion

*Is EC hydrogen native or a product of terrestrial weathering?*

The primary difficulty with measuring the hydrogen content of a meteorite is ensuring that the hydrogen is not a product of terrestrial weathering. This is especially pertinent in ECs, where rare sulphide phases such as oldhamite react readily with atmospheric moisture to produce hydrated sulfates (Avril et al., 2021; Bland et al., 2006; Okada et al., 1981). While we avoided areas with extensive visible terrestrial weathering products, such as rust, much of the remaining meteorite could still have been modified by terrestrial weathering. Indeed, it has been suggested that previous measurements of bulk EC hydrogen concentrations were substantially influenced by the presence of terrestrial water/$OH^-$ contained in weathering products (Peterson et al., 2023).

Figure 1 shows a peak at 2483.4 eV, attributed to the presence of oxidised, $S^{6+}$-bearing species. However, the pre-terrestrial minerals observed petrographically in pristine ECs (including niningerite, oldhamite, and Si-bearing metal) argue these meteorites formed under particularly reducing conditions (Weisberg & Kimura, 2012) that will have led to the total absence of $S^{6+}$-bearing minerals (Klimm et al., 2012). As such, the presence of $S^{6+}$-bearing phases is most easily explained by terrestrial weathering, most likely of native sulfides. This is reinforced by spot spectra measured from a rust-bearing crack in the meteorite, which contained the highest $S^{6+}$ peak amplitudes we measured (Fig. S3). Additionally, a grain of native troilite that has another crack running through it has been transformed into pyrrhotite via reactions with terrestrial water (Fig. S4). Crucially, these two crack spectra contain either no resolvable H-S peak at ~2473.2 eV (the crack containing terrestrial pyrrhotite; Fig. S4) or one of very small amplitude relative to the $S^{6+}$ peak (Fig. S3). Together, these observations suggest that H-S is not a weathering product, so is instead native to EH3 meteorites.

Nonetheless, we chose to still filter the points we analysed to include only those that experienced the lowest extents of terrestrial weathering. We monitored this extent of weathering in each spectrum primarily using the ratio of the amplitude of the $S^{6+}$ peak to that of the FeS peak; as terrestrial weathering converted sulfide to sulfate (Bland et al., 2006) this ratio will have increased (Fig. S5). This ratio varied widely throughout our mapped area, demonstrating that the local extent of terrestrial weathering was strongly heterogenous and was very low in some places (Fig. S6). To minimise the impact of weathering, we chose to only include spectra with an $S^{6+}$:FeS peak amplitude ratio <0.5 in our analysis (Fig. S5), which we consider to be the most pristine (see SI).

*Abundance of H*
Our analysis demonstrates that native H-S exists in fine matrix and mesostasis in ECs, and is in highest concentrations in fine matrix. These observations support previous studies that observed a relationship between S and H concentrations in mesostasis as well as the S-H bond in this phase using Raman spectroscopy (Thomassin et al., 2023). As such, S-H appears to be widespread in ECs.

Our S-XANES analysis suggests that fine matrix accounts for an average of ~5 times as much H-S as mesostasis in LAR 12252. This finding argues that fine matrix is likely the dominant source of H in EH3 chondrites and H-S is likely the dominant form of H in these meteorites. Combined with the observation of the S-H bond in mesostasis alongside previous measurements of the amount of H in organic matter in ECs (Piani et al., 2012), our observation of significant quantities of H in fine matrix mean it is likely that ECs carried more H than had been previously considered based on their petrology (e.g. Marty, 2012). This finding supports recent studies that find relatively high bulk H concentrations in EH3 chondrites (Piani et al., 2020). In all, EH3 chondrites appear to contain higher concentrations of native H, in the form of H-S, than previously thought, which is possibly capable of explaining the entire water budget of the present-day Earth (Piani et al., 2020).

*Source of H*
Despite appearing to host significant amounts of hydrogen in ECs, the phase that carries the H-S bond we identified, and how it formed, were previously unknown. Investigating this further, we found that spectra with large H-S peak amplitudes also exhibited large Fe-S peak amplitudes. In fact, the ratio of these two peak amplitudes varies between only ~0 – 1 (Figure 3) despite absolute amplitudes of these peaks, and relative amplitudes between other pairs of peaks, varying by several orders of magnitude throughout our mapped area (Fig. S2). This relationship suggests that the concentration of H-S is linked to the abundance of iron sulfides.

This variation in relative peak amplitudes can be used to further explore the origin of hydrogen in ECs. We found that spectra with a resolvable H-S peak (i.e., ratio of H-S peak amplitude to iron sulfide peak amplitude >0) tend to contain an iron sulfide peak at lower energies than those that do not contain a resolvable H-S peak (Figure 3a). The energy of the iron sulfide peak reflects the stoichiometry of this phase, with the peak shifting the lower energy as the mineral become more iron deficient (i.e., as the mineralogy evolves from troilite to pyrrhotite; Head et al., 2018; fig S4). This observation suggests that hydrogen is associated with pyrrhotite in ECs.

We can explore this in more detail by examining the relative peak amplitude of each individual spectrum. We find a trend between the energy position of each sulfide peak and the relative peak amplitude of the H-S peak (Figures 3b, S1). This relationship further indicates that the abundance of the H-S bond increases gradually as the amount of pyrrhotite increases. Together, these observations argue that pyrrhotite is key to the source of H.

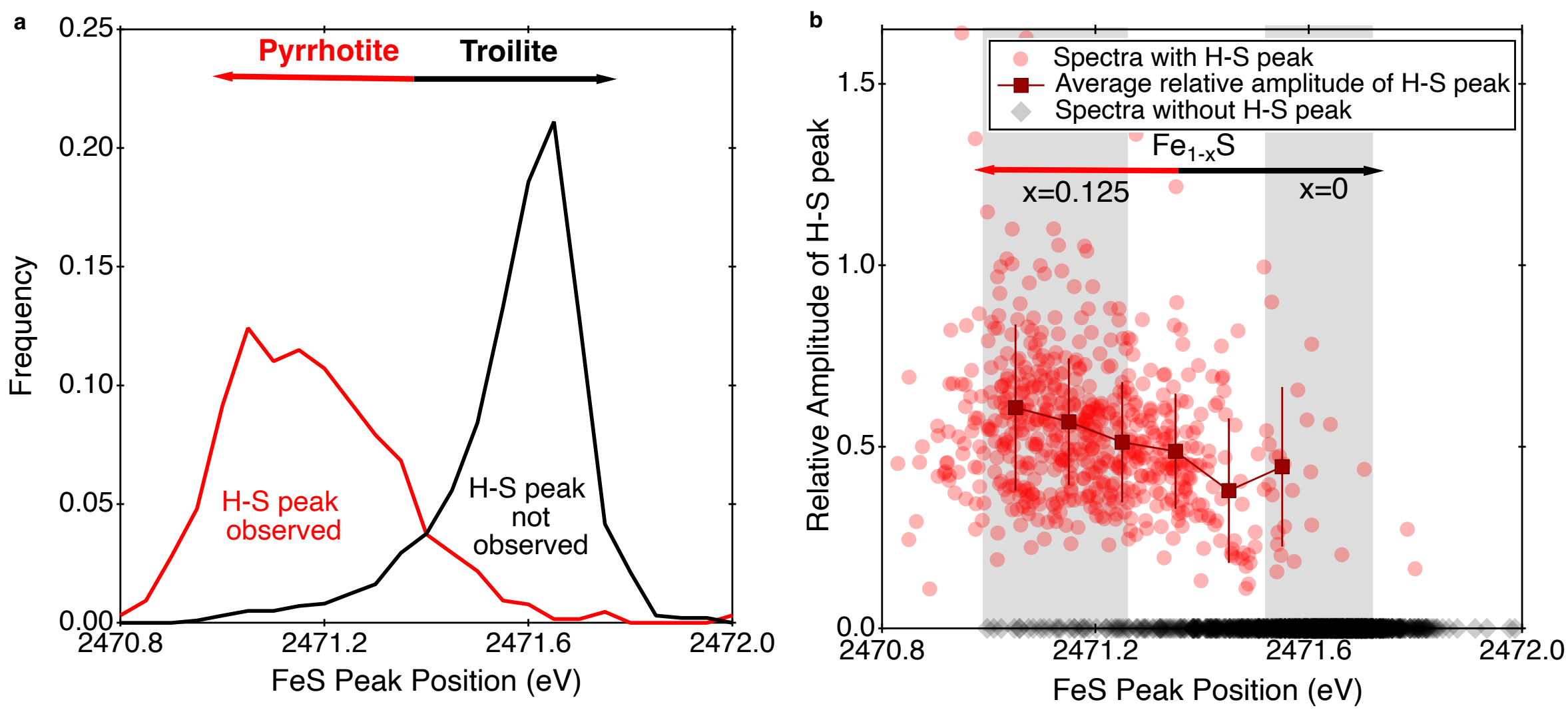

**Figure 3a** The distribution of the position of the FeS peak in all points throughout the mapped area for spectra with (red) and without (black) resolvable H-S peaks. Spectra with energies of the iron sulfide peak that indicate pyrrhotite is the dominant sulfide are more likely to carry the H-S peak, while spectra with iron sulfide peak energies consistent with troilite are less likely to display an H-S peak. **b** The amplitude of the H-S peak normalised to the amplitude of the corresponding FeS peak from each spectrum, plotted against the peak position of the FeS peak, from points included in final analysis. Grey bars indicate the peak location of endmember troilite and pyrrhotite. The transition between these bars represents a change in x in the pyrrhotite chemical formula, $Fe_{1-x}S$, $0<x<0.125$. Black diamonds on the horizontal axis represent spectra with no resolvable H-S peak. The dark red squares represent averages of the individual points in 0.1 eV-wide bands. The relative amplitude of the H-S peak increases as the abundance of pyrrhotite increases, suggesting pyrrhotite is linked to hydrogen in ECs. The slight increase at higher energy reflects the low number of points in this energy region containing an H-S peak, and difficulties identifying the H-S peak when its amplitude is small.

The presence of pyrrhotite is perhaps unexpected given the very reducing nebular environment in which ECs formed (Schrader et al., 2021; Weisberg & Kimura, 2012). Previous measurements of the concentrations of iron sulfides in ECs made using electron probe microanalysis find near-stoichiometric compositions in iron sulfides (Piani et al., 2016; Schrader et al., 2021). Our S-XANES spectra support this result, with the spectra from the centre of macroscopic ($\gtrsim$10 μm) iron sulfides in our map matching reference spectra of troilite (Fig. 1a). However, in the fine matrix and approaching the edges of some macroscopic sulfide grains, the energy of the Fe-S peak shifts to lower values (Fig. S7). This shift argues that pyrrhotite exists as micron- to sub-micron-scale grains in fine matrix and micron- to sub-micron-scale rims on larger troilite grains. Although nebular troilite is more common, nebular pyrrhotite has been observed previously in multiple submicron settings: as rims ~50 nm wide coating larger grains of troilite in carbonaceous chondrites (Harries & Langenhorst, 2013); in the matrix of ECs as submicron grains (Lehner et al., 2014); and as <50 nm grains in anhydrous interplanetary dust particles (Zolensky & Thomas, 1995). Combined with our results, these observations argue that iron sulphides exist as pyrrhotite when they are submicron in scale. This pyrrhotite has been proposed to have formed through low temperature sulfidation of pre-existing troilite in a high $fS_2$ environment in the disk (Harries & Langenhorst, 2013; Kerridge, 1976; Zolensky & Thomas, 1995), or alternatively through the sulfidation of Fe metal by $H_2S$ (Gainsforth et al., 2017). Our results suggest this former process occurred only on the outer part (perhaps <1 μm) of troilite grains, which would cause an entire troilite grain to transform to pyrrhotite if it was below this critical size.

While our analysis indicates that points containing more submicron pyrrhotite are also rich in H-S, the coarse X-ray spot size (5 μm) means we cannot observe whether submicron pyrrhotite itself is the carrier phase. If pyrrhotite were the carrier of this hydrogen, we might expect $Fe_7S_8H_2$ to be present, where iron vacancies are filled by two hydrogen atoms (Cabedo et al., 2024). This phase would be 0.31 wt% H, which would correspond to lower enrichment factors than we measure in points in the fine matrix (Fig. 2b). This difference argues that hydrogen sequestered directly in the structure of pyrrhotite cannot alone explain the concentrations of H-S measured in fine matrix, and another H bearing phase must be present.

Pyrrhotite in ECs is often enclosed in amorphous $SiO_2$-rich material (Lehner et al., 2014), a phase known to store hydrogen bonded to sulfur in these meteorites (Thomassin et al., 2023). Such glassy material could have formed if grains of pyrrhotite accumulated dust particles when they were free-floating in the disk and these assemblages experienced flash heating events. As demonstrated by Cabedo et al., (2024), the elevated temperatures during these events would have led to $H_2S$ being produced through the sacrificial catalytic reaction of pyrrhotite with nebular $H_2$. Due to their adjoining nature, this $H_2S$ could have dissolved into the molten silicate, where it would become trapped upon glass formation during rapid cooling. As such, we propose that H is trapped in the form of $H_2S$ in glassy silicate material that encases grains of pyrrhotite (Lehner et al., 2014). This would mean that the peak we measure in our S-XANES spectra is due to $H_2S$ trapped in silicate glass, as has been observed previously (e.g., Klimm et al., 2012). This process can be summarised through the following reactions, and Fig 4:

$$(1)\ 7FeS + \frac{1}{2} S_{2(g)} \xrightarrow[low\ T]{} Fe_7S_8$$

$$(2)\ Fe_7S_8 + H_{2(g)} \xrightarrow[high\ T]{} 7FeS + H_2S$$

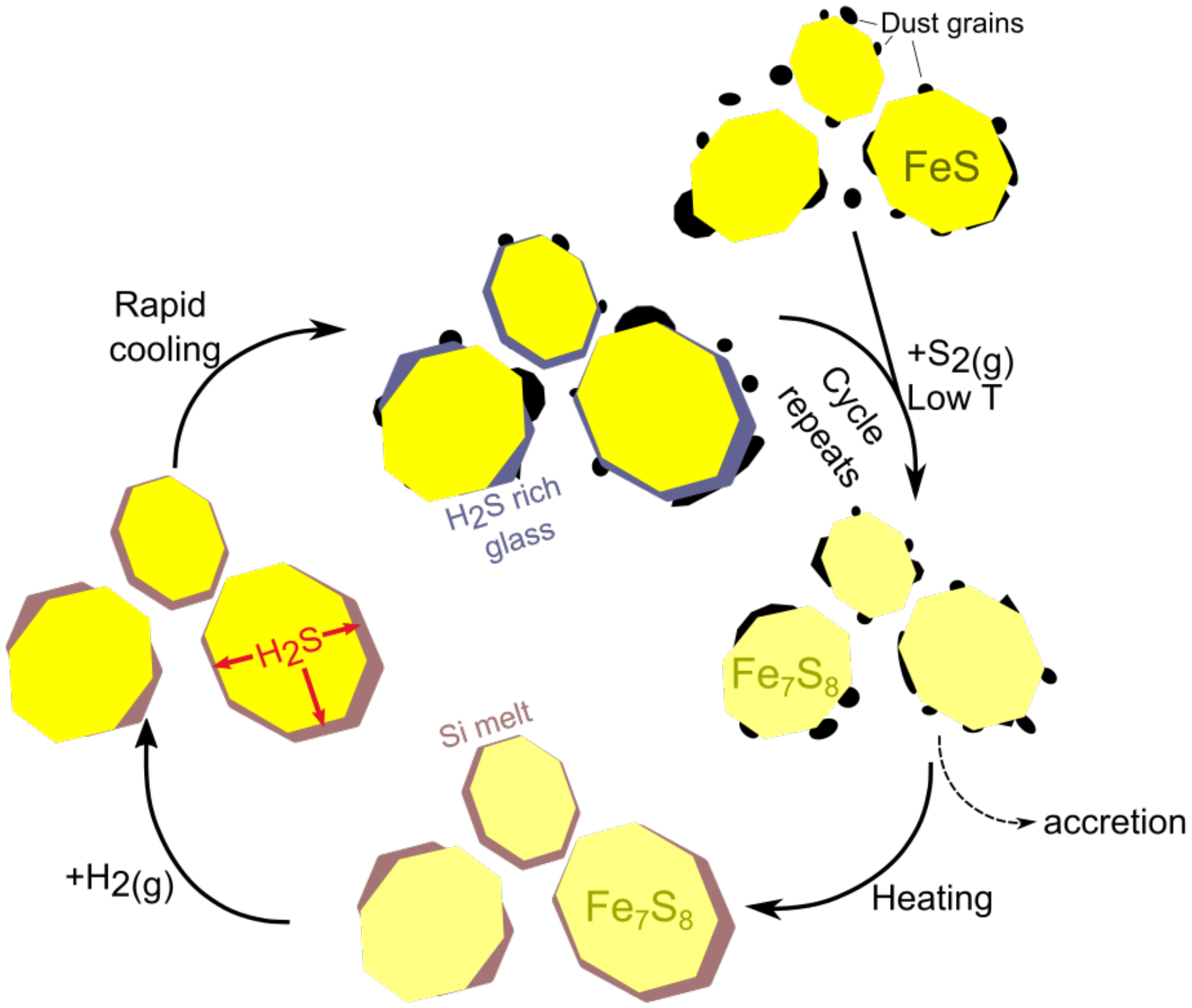

**Figure 4** Schematic outlining the process through which H could be sequestered into ECs. Firstly, submicron FeS grains that become coated in dust particles in the solar nebula are transformed into pyrrhotite at low temperatures in a high $fS_2$ environment (e.g. *Zolensky & Thomas, 1995*). Next, during flash heating events, a sacrificial catalysis reaction with $H_2$ at produces $H_2S$ (*Cabedo et al., 2024*), which is trapped in silicate melt that sits around the grains of sulphide formed when the dust grains melt. Finally, this assembles rapidly cools, produce sulfide grains that are coated in silicate glass, which hosts trapped $H_2S$. This process can repeat, with further dust accumulating onto the surface of the Si-rich glass after it first forms. During/following low temperature sulfidation, accretion can occur to form planetesimals. This cycle results in pyrrhotite existing on a sub-micron length scale surrounded by H-S bearing silicate glassy material, consistent with our S-XANES measurements.

Our data also indicate that hydrogen in chondrule mesostasis and fine matrix are stored in similar ways: predominantly within $SiO_2$-rich glass either contain or are associated with sub-micron sulfide grains. In this case, the carrier phase is the same across both mesostasis and matrix, and we therefore chose not to apply a correction for the phase that carries hydrogen when calculating relative hydrogen peak amplitudes in the matrix compared to mesostasis.

## Implications

In conventional models of the origin of volatile elements on Earth, water was the result of a stochastic, late bombardment of carbonaceous material onto our planet (Alexander et al., 2012). This scenario is capable of explaining Earth's water budget, but means that water and habitable conditions on Earth are a result of a chance scenario. The combined D and $^{15}N$ compositions of the oceans and atmosphere compared to the mantle (Piani et al., 2020) indicate that there is a contribution from this carbonaceous source, at least on the surface of our planet. However, the D and $^{15}N$ isotopic composition of the mantle have been used to argue that this bombardment struggles to explain the bulk of the hydrogen within our planet (Piani et al., 2020). As such, volatiles delivered by bombardment likely correspond to a small proportion of the total budget throughout Earth.

Our findings suggest that H-S in fine matrix in EH3 chondrites is native and potentially the dominant source of H in these meteorites, and that this phase may account for the majority of their bulk H concentrations. These findings indicate that pristine ECs could contain enough hydrogen to explain the entire water budget of our planet, requiring that less water needs to have been delivered to Earth after the main stage of its formation. In this scenario, ECs seemingly became enriched in hydrogen through sulfidation of submicron troilite in an $S_2$-rich nebular gas to form pyrrhotite, which then catalysed a reaction producing $H_2S$ that became trapped in adjoining silicate-rich glass. As such, the water budget of Earth does not need to be the result of a chance bombardment. Instead, it can be a natural and determined consequence of the formation of EC-like material in the hot, reducing, and sulfidising environment of the inner solar system. Additionally, this means that other terrestrial planets that have been proposed to have formed from similar material and in similar environments – including Mars (Brasser et al., 2018), Mercury (Namur et al., 2016), and perhaps Venus – could have had similar early volatile budgets to the proto-Earth as a result of their primary feedstock. A key question now is the hydrogen content of metamorphosed ECs (e.g., EH4-EH6) and aubrites, as well as the proportions of each of these types of meteorites in the proto-Earth.

## Acknowledgements

The authors would like to thank Diamond Light Source for beamtimes (proposals SP31591-1 and SP35606), and staff of beamline I18 for assistance with data collection. The authors would also like to thank the Antarctic Search for Meteorites (ANSMET) program (which has been funded by NSF and NASA and curated by the Department of Mineral Sciences of the Smithsonian Institution and Astromaterials Curation Office at NASA Johnson Space Center) for use of their meteorite samples. JFJB acknowledges funding through UKRI grant number EP/Y014375/1. TJB acknowledges funding through STFC.

## The Source of Hydrogen in Earth's Building Blocks - Supporting Information

Thomas J. Barrett[1], James F.J. Bryson[1], Kalotina Geraki[2]

1. Department of Earth Sciences, University of Oxford, Oxford, OX1 3AN, UK.
2. Diamond Light Source, Harwell Science and Innovation Campus, Didcot, OX11 0DE, UK.

**Email**: thomas.barrett@st-annes.ox.ac.uk

### Methods

### Gaussian Peak Fitting

The energy and shape of a peak in an S-XANES spectrum are controlled by: the redox state of sulfur; the cation; the bond character; and the local bonding environment (Anzures et al., 2020). Unlike more oxidised sulfur compounds, sulfides ($S^{2-}$) with differing cations display characteristic spectra with peaks at specific energies and with distinct shapes (Fleet, 2005). As such, species such as H-S produce a characteristic peak within an S-XANES spectrum. The amplitude of an absorption peak is proportional to species abundance (Fleet, 2005), allowing quantitative estimates of the concentrations of different sulfur species within complex geological samples, such as meteorites, to be recovered.

In order to extract peak amplitudes corresponding to iron sulfide and H-S from our spectra collected from LAR 12252, we fitted 3 Gaussian curves simultaneously across the energy range 2468 – 2478 eV in all 9,024 of our spectra. The first 2 Gaussians have approximately equal width (~1 eV) and were fit to iron sulfide at 2471.2 – 2471.8 eV (depending on stoichiometry) and H-S at ~2473.2 eV. A third curve of larger width (~4 – 5 eV) was fit around 2478 eV, which captures both the second absorption peak of iron sulfide and the step at the S K edge. Although a Gaussian only approximates the shape of this wider feature, we do not use any of the properties of this peak in our analysis, so the quality of fit in this part of the spectra is less crucial. This three-peak approach produced very small fitting errors against the measured spectrum (Fig. 1b in the main text) around the energies of the FeS and H-S peaks (i.e., the key peaks of interest). Gaussian curves fitted to FeS and H-S peaks were constrained by energy (eV) boundaries 2470.8 < FeS < 2472 and 2472 < H-S < 2474, as well as maximum (1.5 eV) and minimum (0 eV) widths, to ensure that fitting was consistent and accurate across all spectra. Certain fits were manually rejected and excluded from calculations where it was obvious that the spectra contained high noise levels or spurious data (white points in Fig. 2b in the main text). High noise levels were most common in spectra with very low sulfur counts (e.g., in the interiors of macroscale metal or enstatite grains).

Differences in monochromator calibration meant that our peaks were not in the locations typically stated in the literature. However, we were able to assign some peaks to certain phases with certainty by measuring spot spectra in large grains (on the scale of hundreds of micrometres to millimetres) that we could identify using optical microscopy. For instance, we were able to unambiguously assign spectra in our region of interest to troilite using this approach (e.g., Fig S4). This led us to identify that all peaks in our data are shifted by very similar energies (~1 eV), compared to literature data, including the database for inorganic sulfur compounds from The European Synchrotron Radiation Facility. This shift allowed us to confidently ascribe the peak we find at ~2473.2 eV to either H-S or elemental sulfur, which are challenging to distinguish due to very similar peak locations (Bose et al., 2024). However, as discussed in the main text, the absence of elemental S in previous micro- and nano-analytical studies e.g. (Lehner et al., 2014; Weisberg & Kimura, 2012) allows us to confidently ascribe this peak to H-S.

### H concentration calculation

In order to calculate the relative concentration of H-S in fine matrix relative to mesostasis, we compared the amplitude of the H-S peak in fine matrix to average amplitude of that in chondrule mesostasis – under the assumption that all H in mesostasis is in the form H-S. For reference, the H concentration of mesostasis in EH3 chondrites has been measured (Piani et al., 2020; Thomassin et al., 2023) across 14 chondrules using background-corrected secondary ion mass spectroscopy, with an average concentration of 708 ± 137 ppm H, but has not been measured in LAR 12252.

We measured S-XANES spot spectra at 15 ($n_1$) individual points within chondrule mesostasis (Fig. S8). We recovered the amplitude of the H-S peak from each of these spectra using our multi-peak fitting method, and calculated the average amplitude, $\overline{Amp}_{Mesostasis}$.

$$\overline{Amp}_{Mesostasis} = \frac{\sum Mesostasis\ peak\ amplitudes}{n_1}$$

We then calculated the normalised peak intensity of H-S at each point throughout our region of matrix relative to mesostasis by normalizing the H-S peak amplitude from each individual point by $\overline{Amp}_{Mesostasis}$.

We identified points in our region of interest that are fine matrix using a series of peak amplitude ratios. We chose to isolate fine matrix for quantification as it represents the dominant phase that stores H-S that we measured. Firstly, we exclude all points with H-S/FeS peak amplitude ratios either equal to 0 or greater than 2. The first of these criteria ensures that points in large (>5 - 10 μm) grains of metal, enstatite, and sulfide (i.e., with no detectable H-S peak) are not included in the analysis. The second of these criteria ensures that we do not include points that have artificially low amounts of iron sulfides due to terrestrial weathering of these phases (see *Weathering* section). Our final step is to consider the amplitude of the H-S peak relative to that found across the range of spectra measured in mesostasis (Fig. S8). By qualitatively assessing our S-XANES map alongside optical microscopy images (Fig. 2 in the main text), we find that fine matrix displays H-S peak amplitudes that are routinely higher than that of mesostasis, reflecting the higher concentration of hydrogen in fine matrix. Therefore, we chose to isolate fine matrix by analysing only points with H-S peak amplitude greater than the range of values measured throughout chondrule mesostasis (Fig S8). Our intention is that all these steps led to only fine matrix being included in our analysis.

We calculated an average, relative hydrogen concentration for fine matrix in our mapped area, $H_{conc}$, by dividing the average H-S peak amplitude in fine matrix by that of the average amplitude in chondrule mesostasis, which produced an enrichment of 9.8 ± 3.1 times. Discussion of the errors associated with this method of quantification can be found *Sources of Error* section.

**Weathering**

Enstatite chondrites are particularly susceptible to weathering on Earth. As discussed in the main text, the detection of an $S^{6+}$ peak in our spectra demonstrates the presence of some, localised terrestrial alteration products throughout our region of interest. The total absence or small relative amplitude of H-S peaks in spectra collected from cracks argues the H-S peak is not created by weathering, and is instead native to the enstatite chondrites. In the case of the rust-bearing crack with a particularly large $S^{6+}$ peak (Fig. S3), we believe that the H-S peak is due to native H that was present before weathering. This because the size of the H-S peak does not scale with the size of the $S^{6+}$ peak. Nonetheless, we chose to exclude points in our analysis that contained evidence of more extensive terrestrial weathering. FeS is thought to react during terrestrial weathering to produce sulphates (Bland et al., 2006). As such, we chose to limit the spectra we analysed to those with a low $S^{6+}$/FeS peak amplitude ratio of <0.5. This threshold ensured regions of the meteorite that experienced more extensive terrestrial weathering were excluded from our analysis, so we are as confident as possible that we are targeting only the most pristine points that may contain pre-terrestrial hydrogen. As discussed in the previous section, we also excluded points with H-S/FeS peak amplitude ratios greater than 2 because this large value could reflect the destruction of iron sulfides due to terrestrial weathering.

Notably, when points with higher ratios of $S^{6+}$/FeS peak amplitude were included in our analysis, the average relative H-S amplitude decreased by 30%. This trend suggests that weathering reactions may actually have decreased the amount of hydrogen stored as H-S. This is consistent with our proposal that H-S is stored predominantly in glassy material surrounding grains of pyrrhotite, because weathering would destroy this metastable phase rather than introducing H. This is not to say that ECs are devoid of hydrogen derived from weathering; rather that our data indicates it is not in the form H-S, and that the majority of the bulk H concentration does not appear to relate to H introduced by terrestrial weathering. Additionally, when peak intensity was measured over time at specific energy (fig. S9), we confirmed that the $S^{6+}$ experienced minor reduction in amplitude and was not being created by the beam.

**Sources of error**

In this study, the dominant source of uncertainty in our relative hydrogen concentration is the natural range of hydrogen concentrations measured in chondrule mesostasis, as well as the unknown hydrogen concentration of LAR 12252 (both bulk and in chondrule mesostasis). We report an H-S enrichment in fine matrix relative to chondrule mesostasis of 9.8 ± 3.1, with the error (2SE) deriving from the spread of H-S peak amplitude values measured within the chondrule mesostasis of LAR 12252 in this study. In both our work (i.e., the spread of H-S peak amplitudes measured among chondrule mesostasis, Fig. S8) and previous studies (i.e., the range of measured H concentrations in mesostasis; Piani et al., 2020; Thomassin et al., 2023), the spread of values in chondrule mesostasis is large, so contributed the most significant source of error. As future studies further constrain the hydrogen content of chondrule mesostasis in LAR 12252, we may be able to extend the conclusions of this work to an absolute hydrogen concentration in fine matrix, and use this to explore a precise, absolute H abundance contributed to Earth's hydrogen budget.

We note that there are additional sources or error. One comes from difficulties in defining and measuring clastic matrix and fine matrix, which have varied definitions in the literature (Lehner et al., 2014; Macke et al., 2010; Rubin et al., 2009). We chose to present the H concentration of fine matrix rather than clastic matrix in the main text because this focusses on the dominant carrier phase of H-S in these meteorites, as well as minimises errors. We also disfavour a clastic matrix approach because we are uncertain that the make-up of our mapped area is representative of clastic matrix; this is particularly true for our mapped area due to the large (~100 μm) chondrule fragment in the centre of our field of view, which will inevitably decrease the H-S concentration of the region owing to its composition being dominantly enstatite.

There are likely to be some errors derived from this approach, particularly relating to the varied definitions of matrix, as well as the coarse beam size (5 μm) making it possible that some points selected as fine matrix include some overlap of other phases. With future studies mapping larger regions we will be able to better quantify the exact contribution of fine matrix H-S to the bulk total.

Additionally, there is potential for error in our quantification due to compositional differences between matrix and mesostasis. Atomic arrangement and the type of phase can influence peak intensity and shape due to various multiple-scattering effects, which can be dealt with using phase correction. However, based on our observation of sub-micron scale sulfides in mesostasis (i.e., a consistent FeS peak), as well as the H-S/FeS peak amplitude ratio being similar between mesostasis and matrix, we propose that the carrier of H-S could also be $SiO_2$-rich glass. Given the expected chemical similarity between $SiO_2$-rich glass in mesostasis as submicron grains in fine matrix, we chose not to apply a correction to account for differences in phases studied.

Finally, there are inevitable sources of analytical error in other aspects of the analysis, such as noise in our spectra and fitting peaks to these data. However, our adopted multi-peak fitting approach minimized these errors as much as was possible, producing only small error values that were typically limited to around 1% of the peak absorption intensity (Fig. 1b). As such, we consider these sources of error as relatively minor.

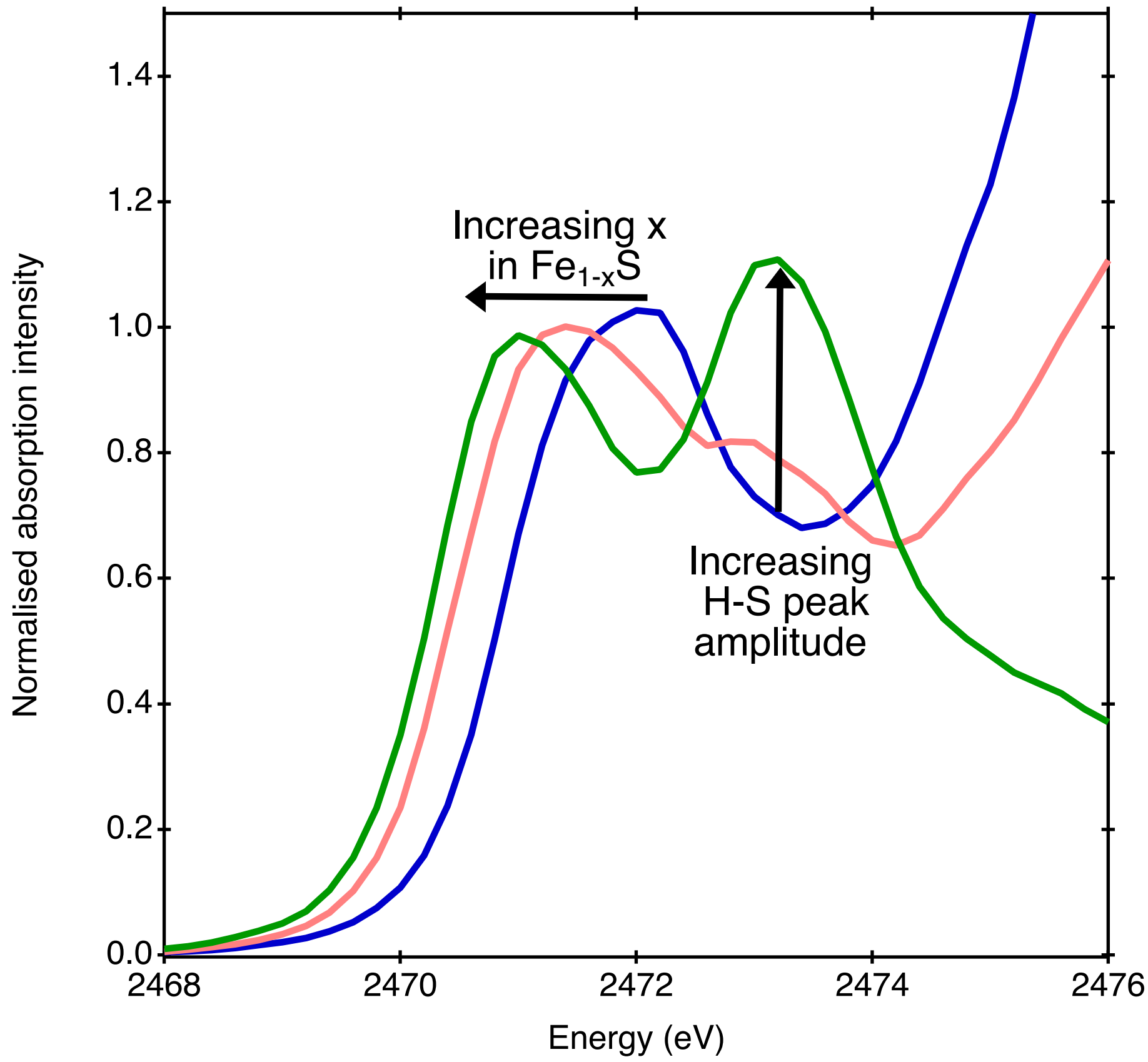

**Fig. S1.** XANES spectra with amplitudes normalized at the FeS peak. The red spectrum is taken from the interior of a grain of troilite (FeS). The green spectrum is taken from the edge of that FeS grain. The blue spectrum is from the fine matrix just outside of this FeS grain. As the spectra move from the interior of a macroscale sulfide to the fine matrix, the position of the iron sulfide peak shifts the lower energy and the relative amplitude of the H-S peak increases. The lateral shift in FeS peak position is indicative of the increase in x in $Fe_{1-x}S$, moving from troilite to pyrrhotite. Together, these trends argue that H is found in regions where pyrrhotite is the dominant iron sulfide. Spectra colour here reflect that of Fig. S7.

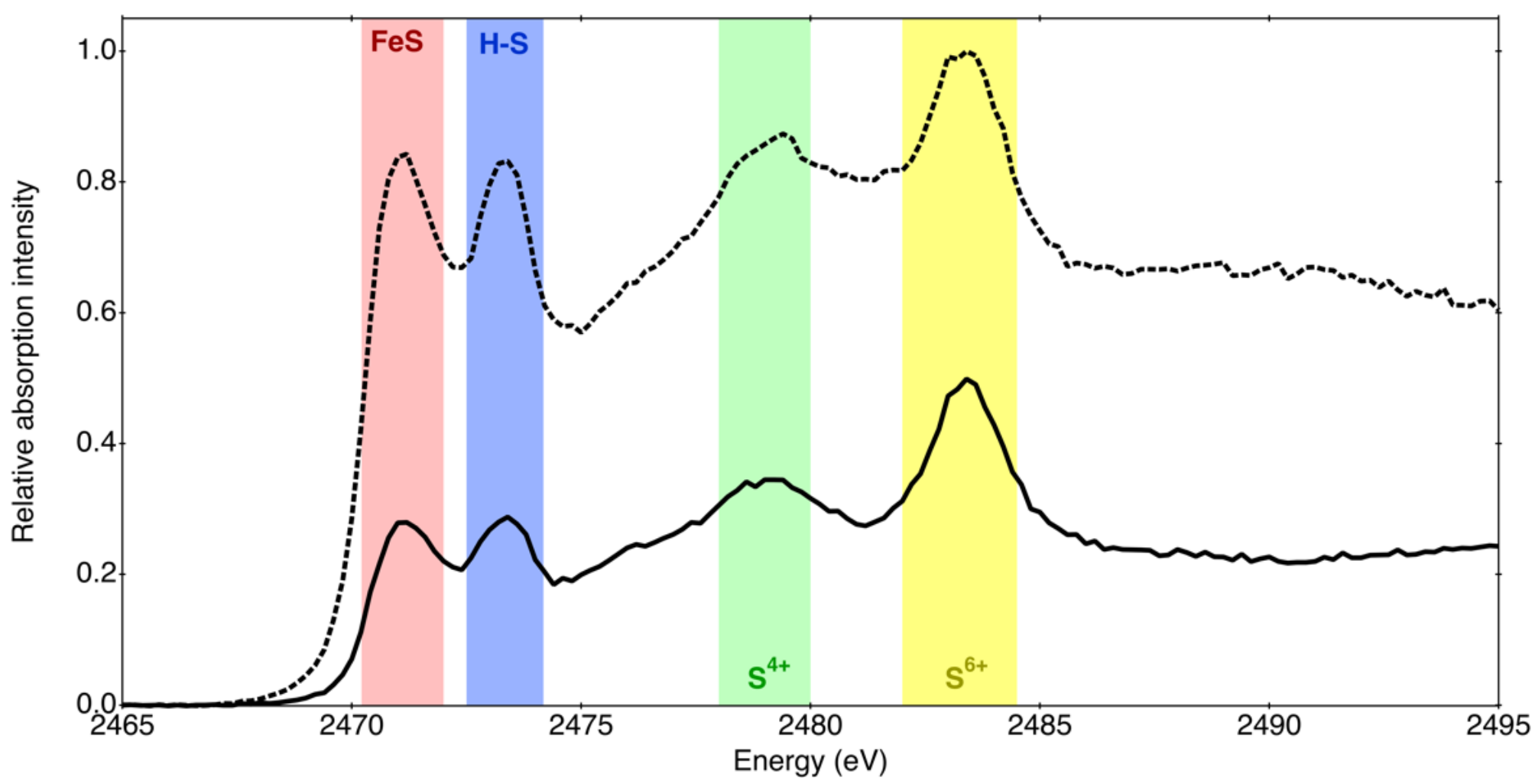

**Fig. S2.** Two spectra from matrix points in LAR 12252, marked by the solid and dashed lines, both normalized to maximum absorption intensity (i.e., at ~2483 eV) in the dashed curve. The ratio between FeS and H-S peak amplitude is similar in these spectra, and others throughout the matrix, demonstrating that H-S concentration is tied to the abundance of iron sulfides. Notably, the absolute amplitude of the peaks varies greatly throughout different areas of matrix in the meteorite, demonstrating the heterogeneity in H-S and iron sulfide abundance.

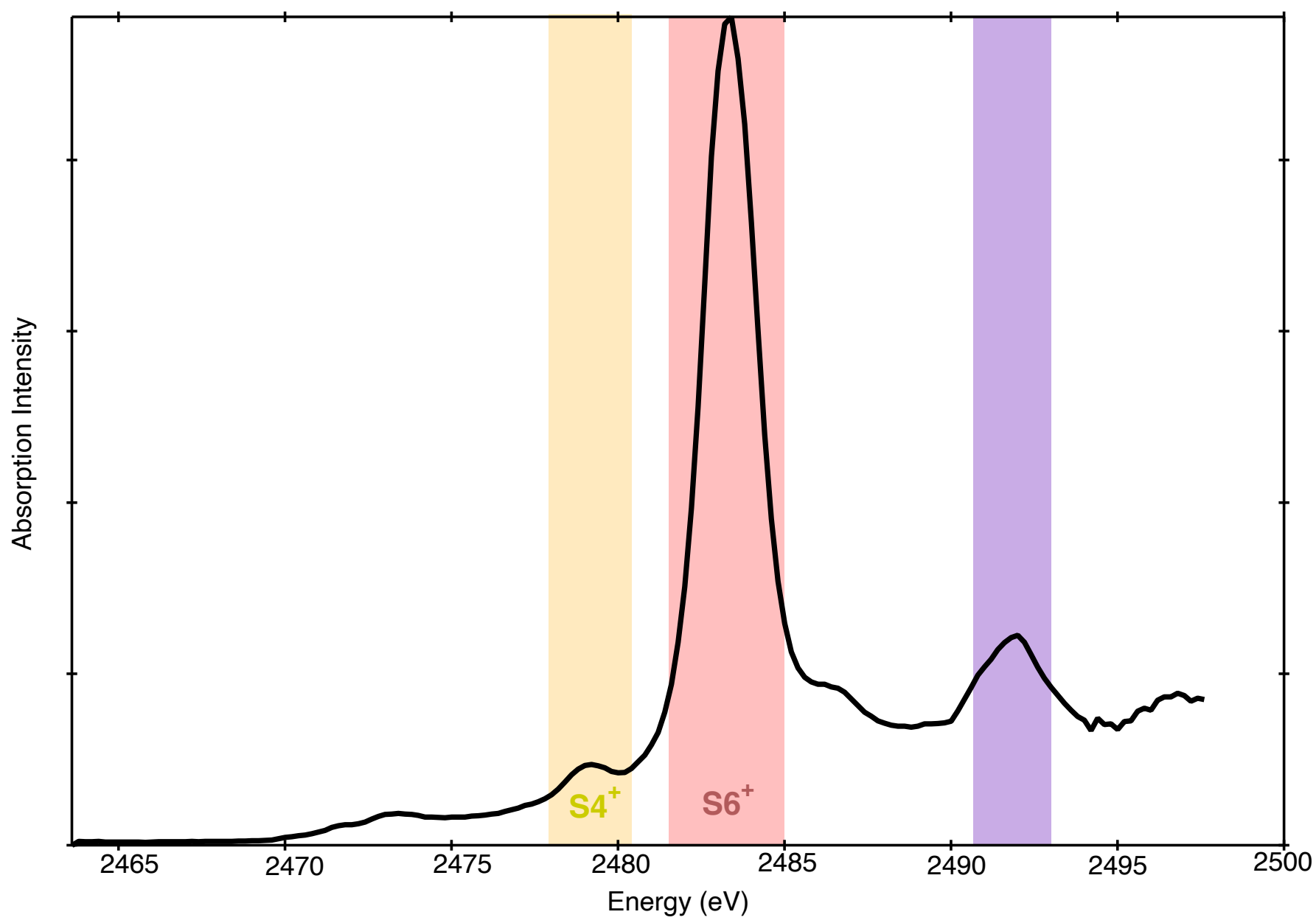

**Fig. S3.** XANES spot spectrum collected from a rust-bearing crack in LAR 12252. The high amplitude peak at ~2483 eV corresponds to the presence of $S^{6+}$-bearing species, which is a terrestrial weathering product. The spectrum contains a peak at 2473.2 eV, though its amplitude is very small relative to the $S^{6+}$ peak, suggesting that H-S is not a weathering product, but instead a small amount of native H-S that survived terrestrial weathering. The purple band corresponds to a secondary absorption peak produced by the sulfate species present here, likely $CaSO_4$ based on spectra from the database for inorganic sulfur compounds from The European Synchrotron Radiation Facility.

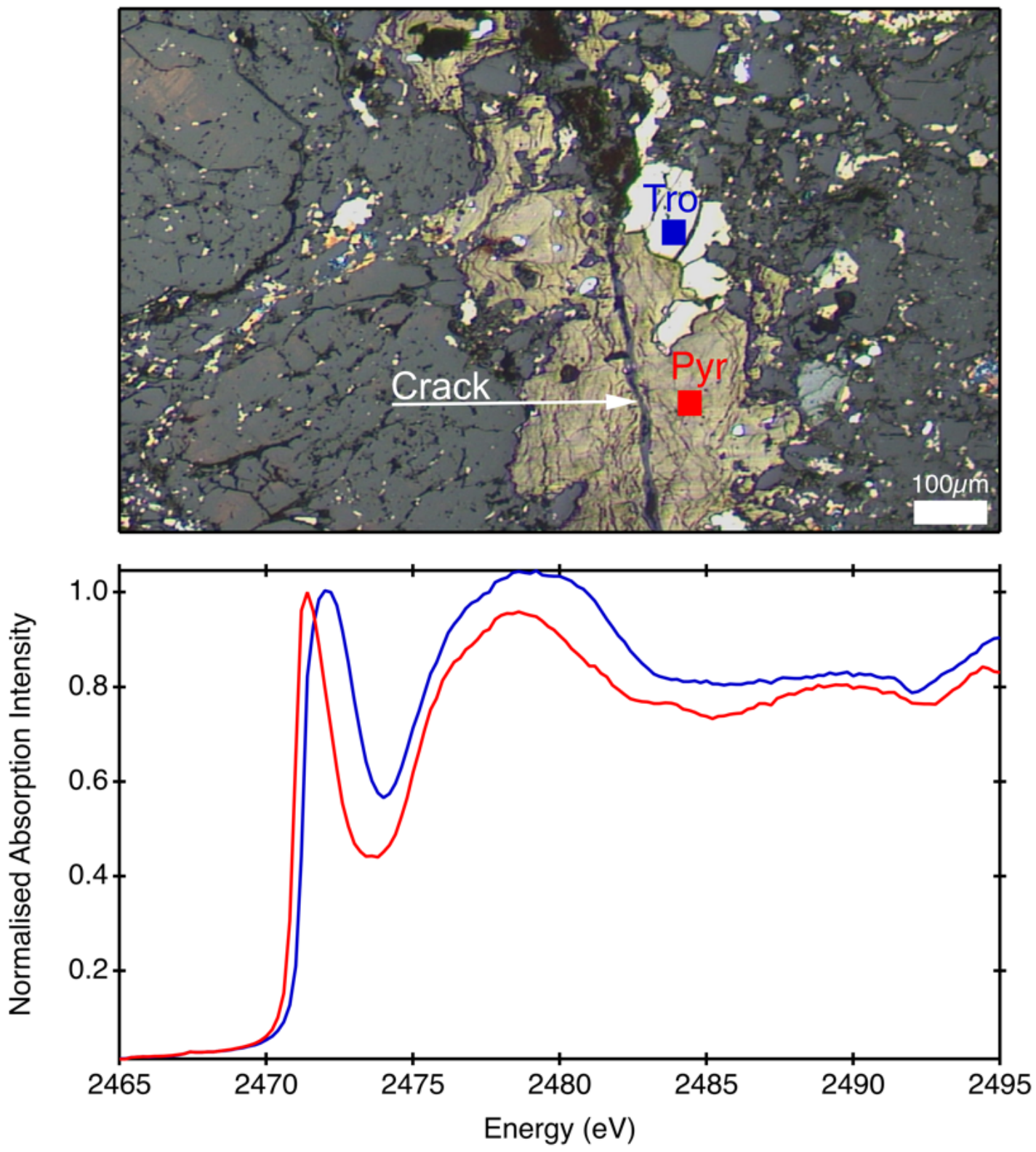

**Fig. S4.** XANES spectra collected from two points within a terrestrially altered crack within the meteorite, with their amplitudes normalized at the initial FeS absorption peak at 2471.2 - 2471.8eV, depending on the stoichiometry. The blue square represents the spot where the blue spectrum was collected, which is a grain of stoichiometric FeS (troilite). The red square is the location where the spectrum in red was collected, which is a grain of pyrrhotite. The pyrrhotite here clearly follows the crack, arguing that it is formed by terrestrial weathering of original troilite. Notably, this terrestrial pyrrhotite does not contain a peak at 2473.4 eV, demonstrating it is devoid of H-S. As such, the creation of pyrrhotite through terrestrial weathering does not lead to the incorporation of H.

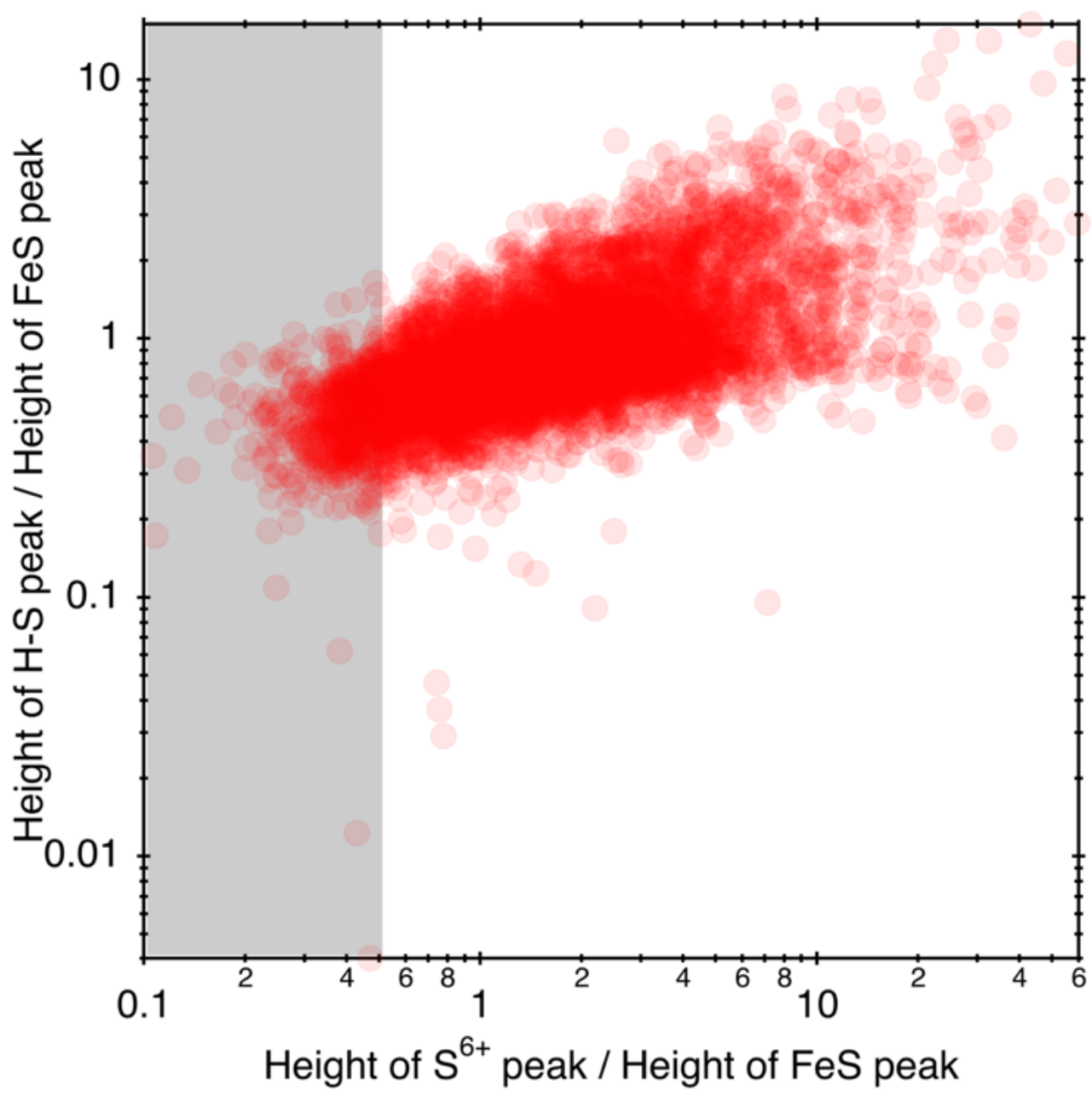

**Fig. S5.** Plot of the amplitude of H-S peak relative to the FeS peak against the amplitude of $S^{6+}$ peak relative to the FeS peak, in spectra containing detectable H-S peaks. The positive trend can be explained by the fact that FeS is the source of both native H-S, locked in the structure of pyrrhotite, and $S^{6+}$ produced in terrestrial weathering

reactions of FeS. To avoid the influence of terrestrial weathering in our analysis, we limited our calculations to points where $S^{6+}$/FeS was below 0.5, as shown by the grey box, which isolated only the most pristine points.

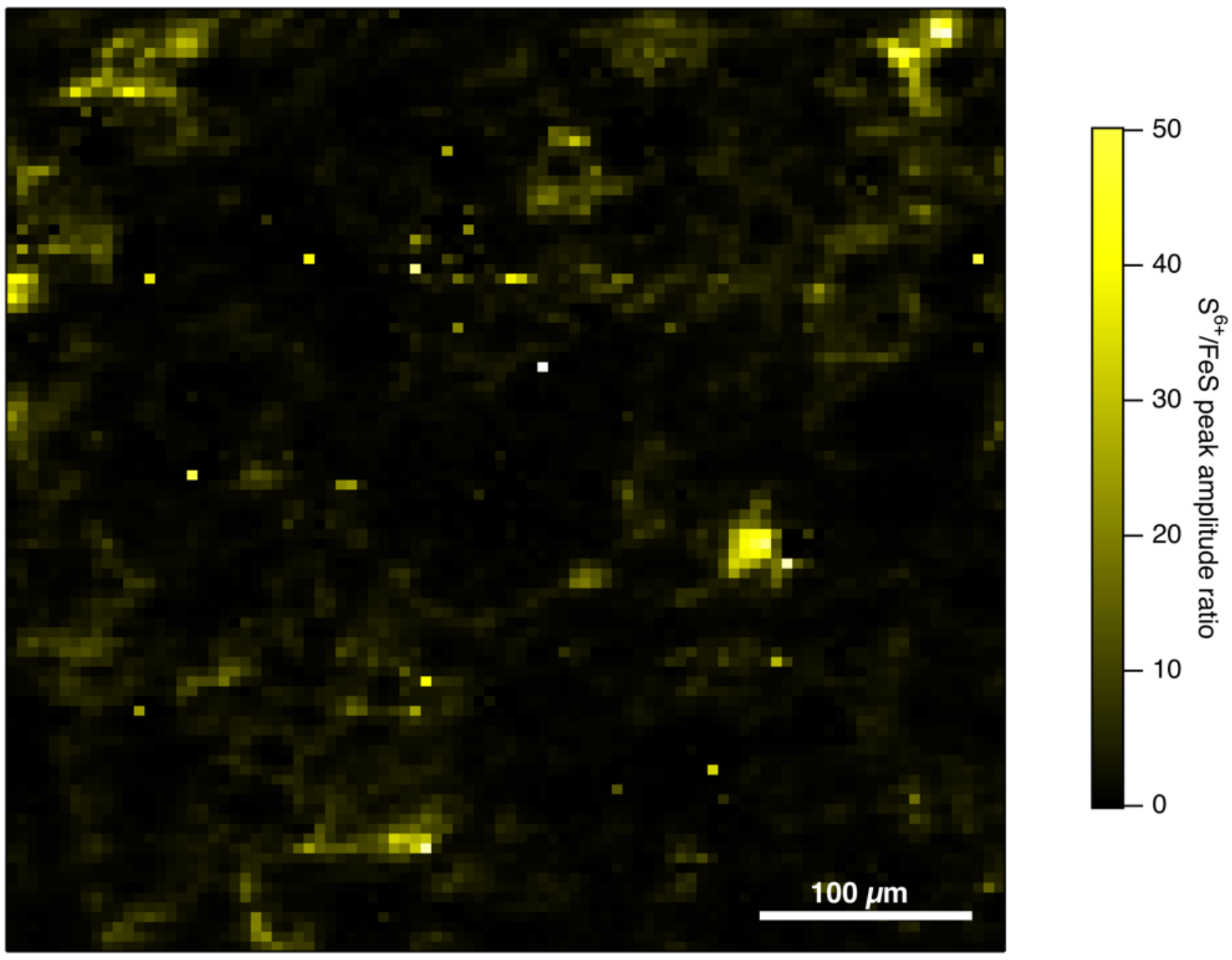

**Fig. S6.** Map of $S^{6+}$/FeS peak amplitude across the mapped area. $S^{6+}$ abundance varies widely and is very heterogeneous, indicating that terrestrial weathering products are mostly in situ at the position of the precursor sulfides. The corresponding optical image is shown in Fig 2a.

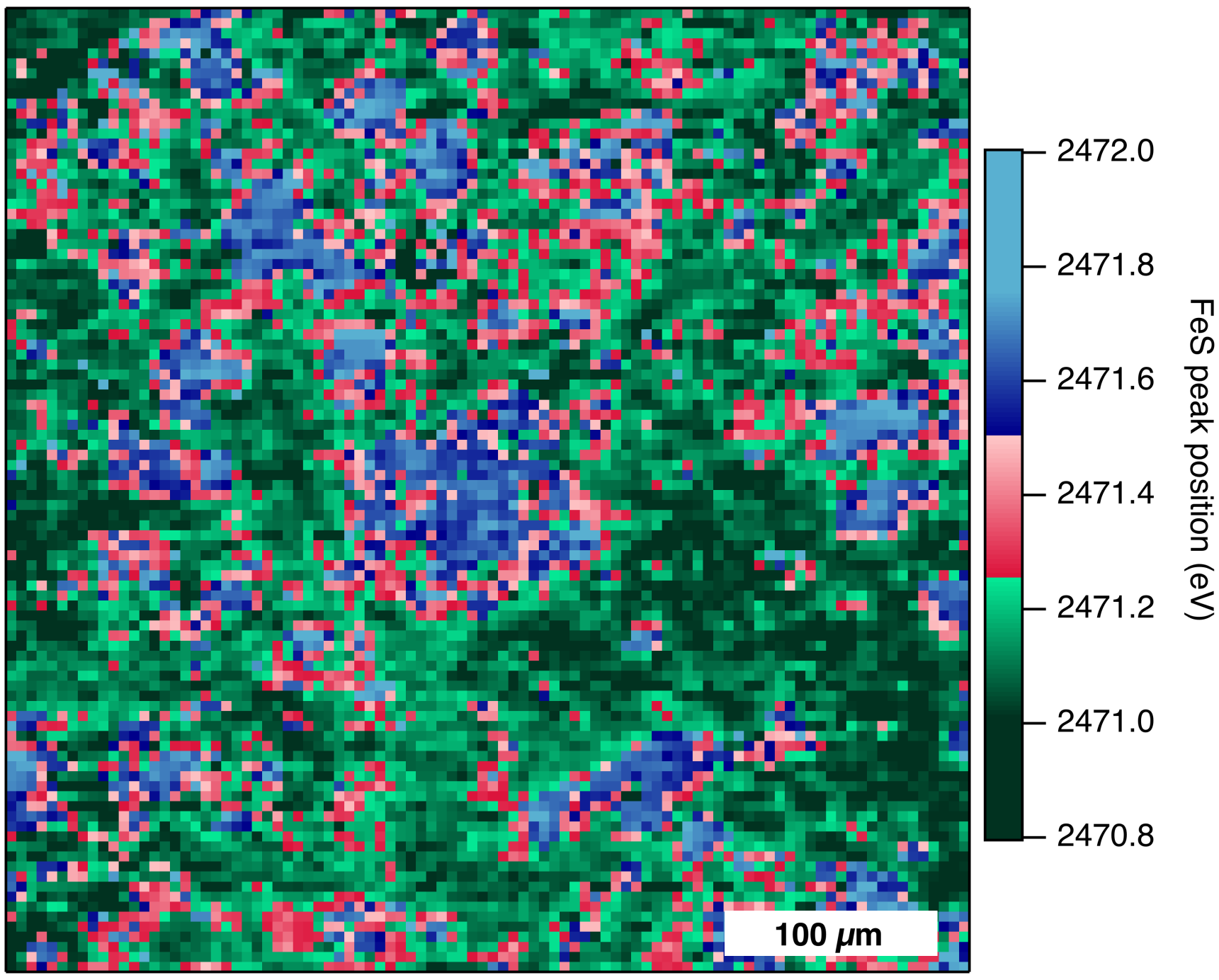

**Fig. S7.** XANES map showing the peak position of the iron sulfide peak throughout our mapped area. In the centre of sulfide grains, the peak is located toward 2472.0 eV (blue colours), indicating the presence of stoichiometric FeS (troilite). At the rims (red colours) of these grains and into the fine matrix material (green colours) , the peak position

shifts to lower energy indicating that iron sulfides are non-stoichiometric (Head et al., 2018). The corresponding optical image is shown in Fig. 2a.

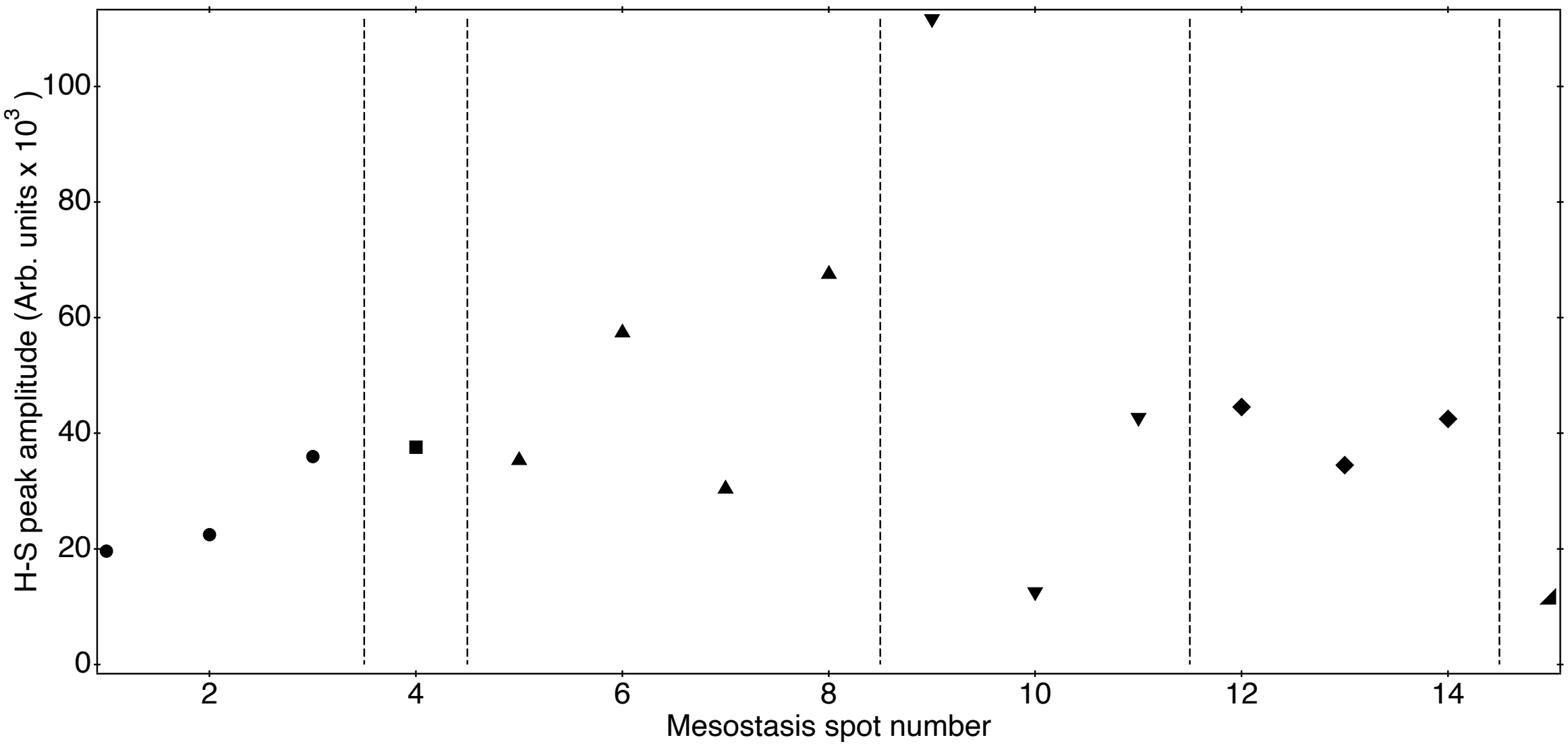

**Fig. S8.** H-S peak amplitudes taken from 15 spot spectra in chondrule mesostasis across 6 different chondrules. Chondrules are distinguished through different shapes, as well as dashed vertical lines. Measurements in each chondrule were taken from different patches of mesostasis within the chondrule.

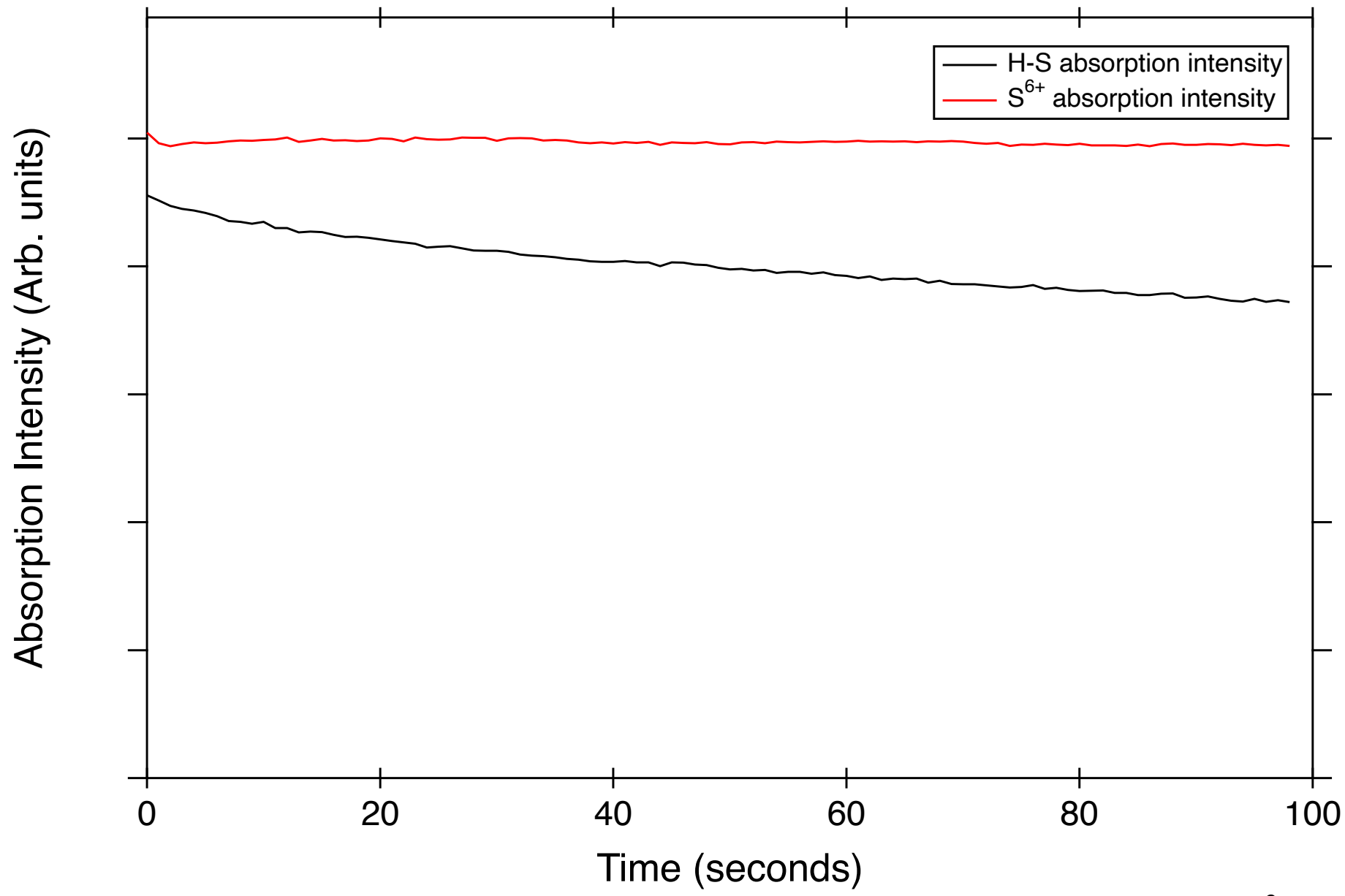

Fig. S9. Absorption intensity over time at the energy of the H-S peak (2473.4 eV), and of the $S^{6+}$ peak (2483.4 eV), measured on a point within the matrix of LAR 12252.

## Supporting References